\documentclass{jaa}

\usepackage{graphicx}
\usepackage{wasysym}
\usepackage{xcolor}
\usepackage[cmr]{emf}

\begin{document}\sloppy

\title{Slow Accretion of Helium Rich Matter onto C-O White Dwarf: \\ Does Composition of WD Matter?}


\author{Harish Kumar\textsuperscript{1}, Abhinav Gupta\textsuperscript{1}
         , Siddharth Savyasachi Malu\textsuperscript{2}, and Shashikant Gupta\textsuperscript{1,*}}
\affilOne{\textsuperscript{1}G D Goenka University Gurugram, India.\\}
\affilTwo{\textsuperscript{2} Indian Institute of Technology Indore, India.\\}


\twocolumn[{

\maketitle

\corres{shashikantgupta.astro@gmail.com}


\begin{abstract}
In binary systems, the helium accretion onto carbon-oxygen (CO) white dwarfs (WDs) plays a vital role in many astrophysical scenarios, especially in supernovae type Ia.  Moreover, ignition density 
for accretion rate $\dot{M} \lesssim 10^{-9} M_{\odot}$ $yr^{-1}$ in helium accreting CO white dwarfs decides the triggering mechanism of a supernova explosion which could be either off-centre helium flash or central carbon flash. 
We aim to study the accretion of helium with a slow accretion rate $5 \times 10^{-10} M_{\odot}$ $yr^{-1}$ 
onto relatively cool and hot white dwarfs of different abundances of carbon and oxygen. 
The simulation code “Modules for Experiments in Stellar Astrophysics” (MESA) has 
been used for our study. We analyze the variation in several properties like surface gravity (g), 
helium luminosity ($L_{He}$), and effective temperature ($T_{eff}$) during the accretion phase of 
the white dwarfs. We also calculate the ignition density (${\rho}_{He}$) and ignition temperature ($T_{He}$) of helium burning. 
As expected, the size of WD decreases and g increases during the accretion. However, 
a red-giant-like expansion is observed after the rapid ignition towards the end. 
The dependence of helium accreting WD evolution 
on its composition has also been explored in this study. We find that white dwarfs of the lower abundance of carbon accrete slightly longer before the onset of helium ignition.
\end{abstract}

\keywords{Supernovae Ia---White Dwarfs---Accretion---Helium fusion---Stellar evolution.}
}]

\doinum{}
\artcitid{\#\#\#\#}
\volnum{000}
\year{2021}
\pgrange{1-5}
\setcounter{page}{1}
\lp{1}

\section{Introduction}

In the binary system, helium accretion onto a Carbon-Oxygen (CO) white dwarf (WD) plays a vital role in many astrophysical scenarios (Peng $\&$ Ott 2010). Most significantly, it may be relevant to the Supernovae Ia (SNe Ia) progenitors (Kuuttila 2021; Hillebrandt $\&$ Niemeyer 2000; Hillebrandt {\em et al.} 2000). The essential evolutionary paths that lead to SN Ia are following: 1) Semi-detached close binary stars in which WD grows its mass up to the Chandrasekhar limit ($M_{Ch}$) by accreting matter directly from the companion (e.g. (Nomoto {\em et al.} 2007; Kasen $\&$ Woosley 2007; Nomoto {\em et al.} 2013; Neunteufel {\em et al.} 2016; Wang {\em et al.} 2017); 2) Explosions of sub-Chandrasekhar mass WD via “double-detonation” mechanism. The He-layer at the surface of an accreting WD triggers detonation via shock waves (e.g. Nomoto 1982; Ropke {\em et al.} 2006). 
In the case of helium accretion onto a WD, Triple-alpha (3$\alpha$) reaction is the key to the nuclear fusion of helium into carbon. Generally, triple-alpha (3$\alpha$) reaction occurs at ($10^8$ K) by the resonant method. However, below $10^8$ K, it can occur via a non-resonant reaction which often takes place at slow accretion rates such as $\dot{M} \lesssim 10^{-9} M_{\odot}$ $yr^{-1}$. The study is crucial in understanding the helium ignition density, which determines the triggering mechanism of a supernova explosion, i.e., the explosion will occur via off-centre helium ignition or central carbon ignition (Nomoto 1982b).
In this paper, we study the effect of carbon abundance of helium accreting WDs with a slow accretion rate on the properties like surface gravity (g), radius (R), effective temperature ($T_{eff}$), and helium luminosity ($L_{He}$). We also determine the ignition density ($\rho_{He}$) and ignition temperature ($ T_{He}$) of helium.

\section{Numerical Method}
\label{sec:method}
We have used “Modules for Experiments in Stellar Astrophysics” (MESA), which is a one-dimensional stellar evolution code (version 12778; Paxton {\em et al.} 2011; Paxton {\em et al.} 2015; Paxton {\em et al.} 2019) to study the evolution of WD during accretion. One dimensional means that the WD is spherically symmetric, and changes in WD structure occur only in the radial direction. To first order, this is an excellent approximation. For our study, we have considered CO WD of mass 0.85 $M_{\odot}$, which lies in the middle order of the mass range of WDs. Two different compositions of WDs with low and high carbon fractions have been considered, i.e., with C and O abundances (0.3, 0.7) and (0.7, 0.3), respectively. The effective temperature of WDs has also been taken in the moderate range. The parameters for the initial models of the C+O white dwarf are given in Table 1. Thus depending on the composition and effective temperature, there are four models of WDs for the study. A slow accretion rate has been chosen for study so that the triple-alpha reaction is non-resonant in all cases.
Various test suits, subroutines, functions, and control options are available in MESA to evolve stars and calculate various parameters. For our purpose, we have used the \texttt{make{\_}co{\_}wd} test suite to produce a WD of a given mass from a zero-age main sequence (ZAMS) star. We then let the WDs cool to certain specific temperatures and finally change the carbon and oxygen abundances artificially using the relax{\_}composition option in the Inlist file. For helium accretion onto the WD, the test suite wd3 has been used with more than 1500 grid points considered for our calculations. The opaqueness of the star material for the radiation is expressed in opacity, which has been taken from OPAL opacity tables. Various nuclear reactions take place during the evolution of a star as well as during the accretion. The subroutine \texttt{co{\_}burn.net} which includes 57 nuclear reactions, has been used for burning isotopes of helium, carbon and oxygen elements. Since the density of matter in the WD is very high, the pycnonuclear reactions and screening effects become important, which have been incorporated using \texttt{set{\_}rate{\_}3a} = 'FL87' (Fushiki $\&$ Lamb 1987). 

\section{Results And Discussion}
\label{sec:result}
\subsection{ Initial and Ignition Parameters}
As discussed in section-2, WDs of mass 0.85 $M_{\astrosun}$, with different abundances of C\&O, have been used for our analysis. The WDs have been generated by evolving a star of initial mass 4 $M_{\odot}$ from ZAMS followed by the main sequence and red giant phase. Then the abundances of C and O of the white dwarfs are controlled artificially by MESA, permitting them to cool to get the desired value of effective temperatures. The evolution in the WD phase is halted for two different values of effective temperature (Teff = 38000 K and 75000 K). Two different compositions of these WDs have been considered, one with 30\% C \& 70\% O, and the other with 70\% C \& 30\% O. Thus, there are four models of WDs shown in Table 1, along with the numerical values of their physical parameters. 
\begin{table}[htb]
\tabularfont
\caption{Initial Parameters of WD}
\label{table:1} 
\begin{tabular}{lcccc}
\topline
         \hline

Parameters & A & B & C& D\\\midline
C Fraction & 0.3 & 0.7 & 0.3 & 0.7\\
$T_{eff}(10^{4}K)$ & 3.7 & 3.7 & 5.2 & 5.2\\
$T_{c}(10^{7}K)$ & 5.4 & 5.5 & 6.1 & 6.1 \\
$\rho_{c}(10^{6}g$ $ cm^{-3})$ & 3.5 & 3.5 & 3.4 & 3.4 \\
Radius (R) $(10^{-3}R_{\odot})$ & 9.7 & 9.8 & 10.3 & 10.4 \\
\hline
\end{tabular}
\end{table}
                                              
The parameters related to helium ignition are presented in table 2, where $T_{He}$ and $\rho_{He}$ are the temperature 
and density of the helium zone of the WD at which helium ignites. $\Delta M_{He}$ is the total mass of helium accreted 
onto WD before helium ignition. $T_C$ and $\rho_C$ are the central temperature and density of the WD at the ignition point, 
which is defined as the point at which the nuclear time scale, $\tauup_{He}=\frac{C_{P}T} {\emf_{He}}$, from helium burning 
is almost equal to $10^6$ years(Nomoto 1982). Here $C_P$ represents the specific heat, and $\emf_{He}$ is the nuclear energy generation 
rate from helium burning.     
The initial parameters related to the WD used in our study are presented in table-1. As discussed earlier, we have 
taken $T_1$ and $T_2$ as the effective temperature of WDs. Two different compositions of the CO WD have been considered, 
i.e., with carbon fraction $0.3$ and $0.7$. These are shown by models A \& B for relatively cool and C \& D for relatively 
higher temperatures. Core temperature and densities have been shown in the table-1. 
The WDs have been subjected to the accretion of helium-rich matter with two different accretion rates. 
One of the accretion rate is as slow as $5 \times 10^{-10}$ $M_{\odot}$ $yr^{-1}$. In this case, the triple-alpha 
reaction takes place via the non-resonant method. The other accretion rate is high ($5 \times 10^{-8}$ $M_{\odot}$ $yr^{-1}$) 
in which triple-alpha reaction occurs via resonance. 

\begin{table}[htb]
\tabularfont
\caption{He Ignition Parameters of WD with $5 \times 10^{-10}$ $M_{\odot}$ $yr^{-1}$}
\label{table:2} 
\begin{tabular}{lcccccc}
\topline
         \hline
Parameters & A & B & C& D\\\midline

t(10$^8$yr) & 9.592 & 9.541& 9.602 & 9.535 \\

$\Delta M_{He}(M_{\odot})$ & 0.48 & 0.477 & 0.48 & 0.4768 \\

T$_{He}$ (10$^{7}$ K) & 4.52 & 4.6 & 5.95 & 6.28 \\
$\rho_{He}(10^{7}g$ $ cm^{-3})$ & 9.97 & 9.4 & 12.6  &11.3 \\
$T_{c}(10^{7}K)$&3.75 & 3.75 & 3.73& 3.73 \\

$\rho_{c}(10^{8}g$ $ cm^{-3})$ & 9.06 & 8.26 & 9.03  & 8.1 \\
\hline
\end{tabular}

\end{table}
\subsection{Variation in Surface Gravity}
 As the WDs accrete helium with a slow accretion rate $5 \times 10^{-10}$ $M_{\odot}$ $yr^{-1}$, a helium zone gradually 
 forms on the C+O core and thus the mass of the white dwarf increases. This rise in mass compresses 
 the matter in the interior to higher densities. The radius of the white dwarf then decreases because 
 of the mass-radius relation of white dwarf $ M \propto R^{-\frac{1}{3}}$ . At the same time, the accreting white dwarf loses 
 energy in the form of radiation and neutrino at a rate of radiative luminosity $L_{ph}$ and neutrino luminosity $L_{\nu}$, 
 respectively. As a result, the cooling of the white dwarf takes place. Variation of surface gravity (g) and 
 radius (R) during accretion is presented in Fig 1. A small portion towards the end of the simulation is 
 zoomed in and is shown in the right panel. A sudden fall in g and a sudden rise in R is visible towards the end.
\begin{figure}[!t]
 \centering \includegraphics[width=1.0\columnwidth]{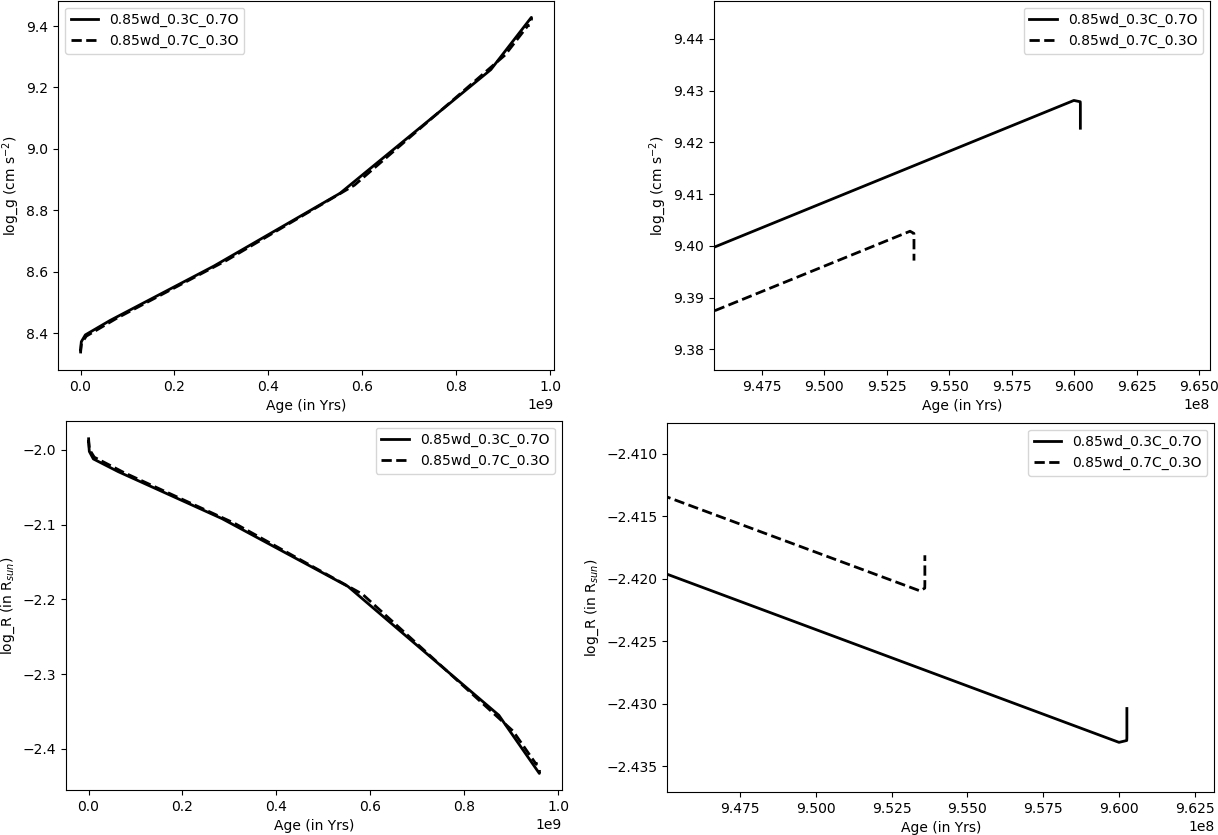} 
  \caption{The grid of plots shows the variation in surface gravity (g) and radius (R) of the 0.85 $M_{\odot}$ hot 
  white dwarf during the age of helium accretion. Top left plot: shows the variation in surface gravity g of 
  the 0.85 $M_{\odot}$ white dwarfs of 0.3 C and 0.7 C carbon abundances; Top right plot: is the variation in surface gravity (g) 
  towards the end of the run. Bottom left plot: gives the variation in radius R (in $R_{\odot}$) of the 0.85 $M_{\odot}$ 
  white dwarfs of 0.3 C and 0.7 C carbon abundances; Bottom right plot: is the variation of radius R towards the end of run.}
\label{fig:1} 
\end{figure}
\begin{figure}[!t]
\centering  \includegraphics[width=1.0\columnwidth]{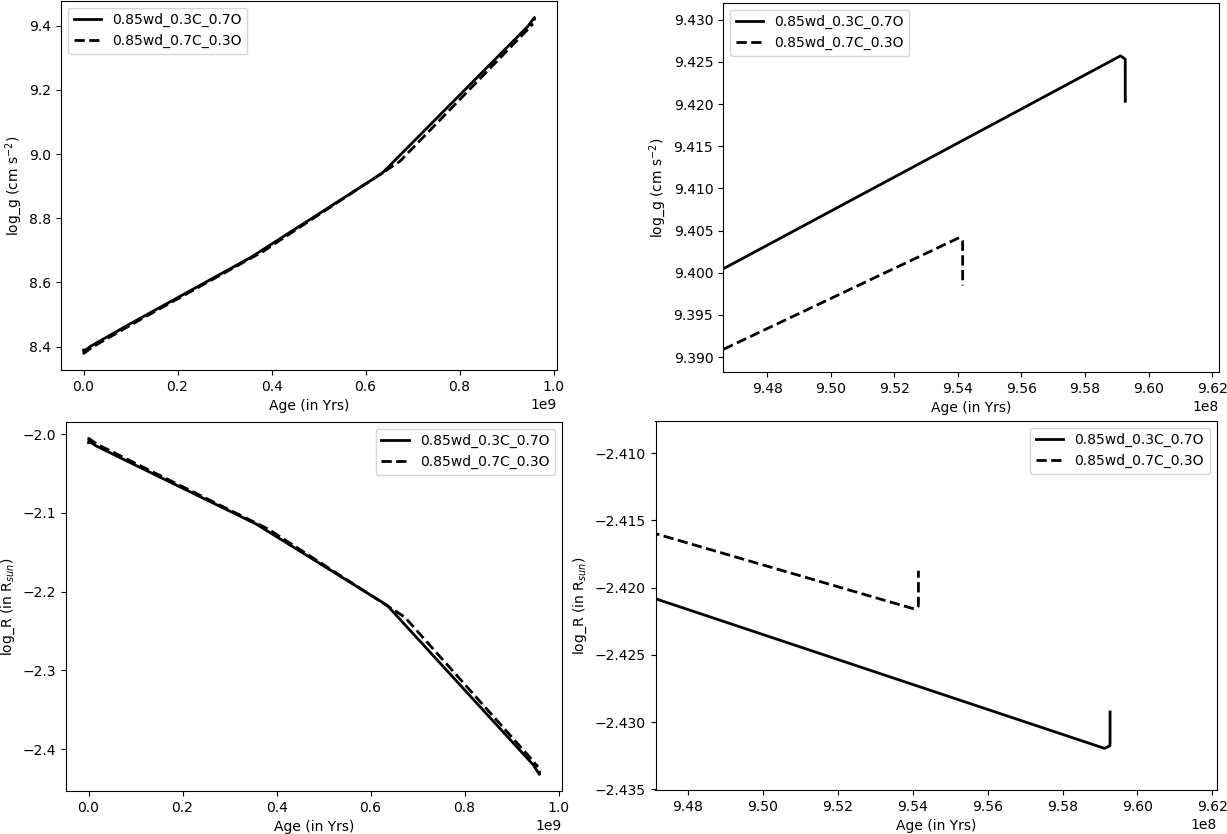} 
\caption{ The grid of plots shows the variation in surface gravity (g) and radius (R) of the 0.85 $M_{\odot}$ relatively 
cool white dwarf during the helium accretion. Top left plot: shows the variation in surface gravity g (in g $cm^{-3}$) of 
the 0.85 $M_{\odot}$ white dwarfs of 0.3 C and 0.7 C carbon abundances; Top right plot: is the variation in surface gravity (g) 
at the end of the run. Bottom left plot: gives the variation in radius R (in $R_{\odot}$) of the 0.85 $M_{\odot}$ white dwarfs 
of 0.3 C and 0.7 C carbon abundances; Bottom right plot: is the variation of radius R at the end of run.}

\label{fig:2}
\end{figure}

\subsection{Variation in Helium Luminosity}
Next, we explore the variation in Helium luminosity ($L_{He}$) of WDs during the accretion phase. 
All the four models (A, B, C and D) have been accreting helium-rich matter at the same rate 
of $5 \times 10^{-8} M_{\odot}$. As mentioned earlier, the carbon and oxygen abundances of 
these WDs are (C, O): (0.3, 0.7) and (0.7, 0.3) by mass. The results are shown in Fig 3 and Fig 4 
for relatively cool ($T_{eff}$ = 38000 K) and hot ($T_{eff}$ = 75000 K) white dwarfs, respectively. 
In the early stage of accretion, the temperature at the bottom of the accreted layer is not enough to 
ignite helium, and hence $L_{He}$ is negligible. The temperature slowly increases. When a sufficient 
amount of matter has been deposited on the WD, the temperature attains a value such that the nuclear time 
scale ($\tauup_{He}$) becomes of the order of $10^6$ years, which is the ignition point of helium. Helium Luminosity 
shoots up at this point. From Table 2 and Fig 3, helium in cases A and B ignite at different time scales, and it is observed 
that helium ignition starts early if the carbon fraction is large compared to the oxygen fraction. However, the difference is tiny. 
The same behaviour is observed for the hot models, as shown in Fig 4 and can be verified from Table 2. It is observed 
that helium luminosity $L_{He}$ increases rapidly at the end of the simulation due to the rapid helium burning at the bottom of 
the accreted helium layer. 
\begin{figure}[!t]
  \centering\includegraphics[width=1.0\columnwidth]{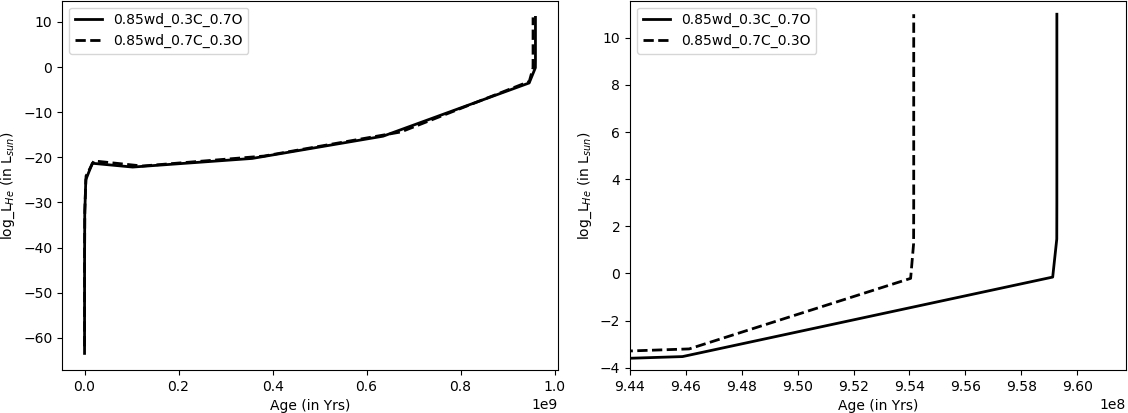} 
\caption{The left panel shows the change in helium luminosity $L_{He}$ (in $L_{\odot}$) of 0.85 $M_{\odot}$ cool white dwarfs of different carbon and oxygen with the age of accretion. The right panel shows the change in helium luminosity $L_{He}$ at the end of the run.}
\label{fig:3}
\end{figure}
\begin{figure}[!t]
  \centering\includegraphics[width=1.0\columnwidth]{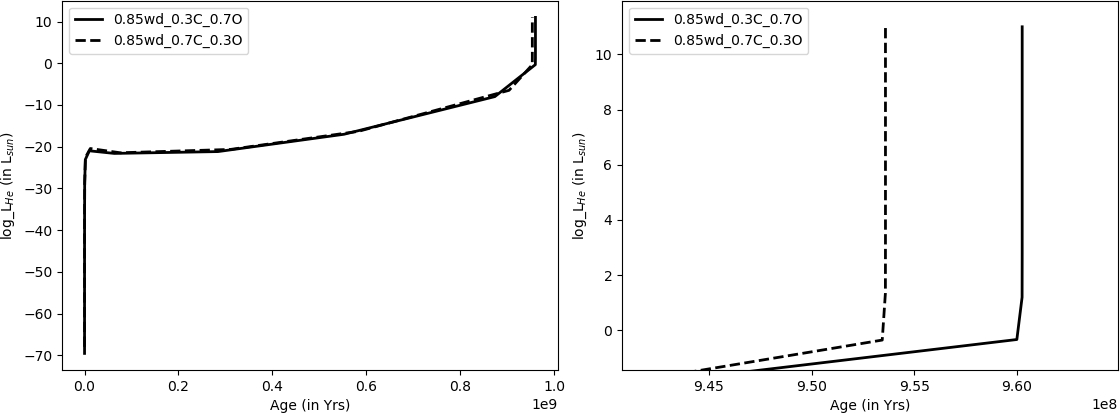} 
\caption{The left panel shows the change in helium luminosity $L_{He}$ (in $L_{\odot}$) of 0.85 $M_{\odot}$ hot white dwarfs of different carbon and oxygen with the age of accretion. And the right panel shows the change in helium luminosity $L_{He}$ at the end of the run.}
\label{fig:4}
\end{figure}

\subsection{Variation in Effective Temperature}
From fig 5, it is observed that the cooling effects dominate during the early phase of accretion for both 
the relatively cool white dwarfs (Cases A and B with effective temperatures 38000 K). In other words, the energy 
losses in the form of radiation and neutrino emission are more significant than the heating effects due to the compression 
of accreted matter. This results in decreasing the effective temperatures at the initial stage. However, after this stage, 
the heating effect dominates over the cooling effect and effective temperature increases. An almost similar trend is observed 
in relatively hot WDs (cases C and D with an effective temperature 75000K) with slight differences (see Figure 6). 
For instance, the effective temperature decreases rapidly in the beginning and after that remains almost constant for 
a while before the rise. During the constant $T_{eff}$ phase, the cooling effect is almost compensated by heat generated due 
to compression of matter. After the constant $T_{eff}$ phase, the WD heats up due to compressional heating, and effective temperatures 
rise for all models. Towards the end, the ignition of helium takes place in the bottom of accreted helium onto C+O white dwarf 
in all models (A to D). One should note from Table 1 that the final temperature of the core is smaller than the initial core 
temperature indicating that the core of WD cools during the accretion. It is to be noted that no significant difference is seen in 
the effective temperature of WDs of different compositions.

\begin{figure}[!t]
  \includegraphics[width=1.0\columnwidth]{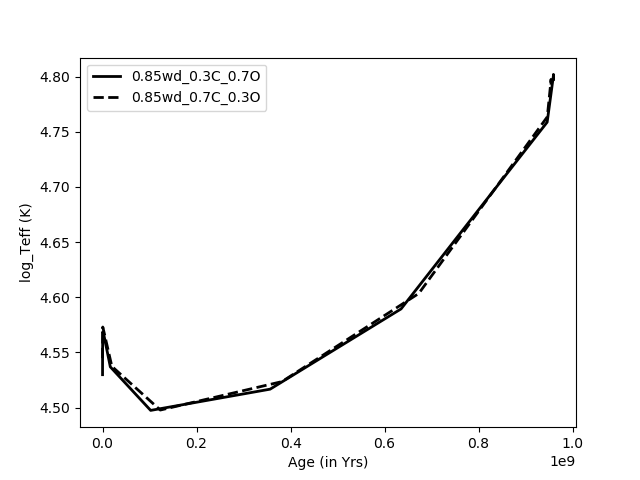} 
\caption{This plot shows the variation in effective temperature $T_{eff}$ with the long term evolution of helium accreting 
0.85 $M_{\odot}$ white dwarf having initial effective temperature 38000 K.}
\label{fig:5}
\end{figure}
\begin{figure}[!t]
  \includegraphics[width=1.0\columnwidth]{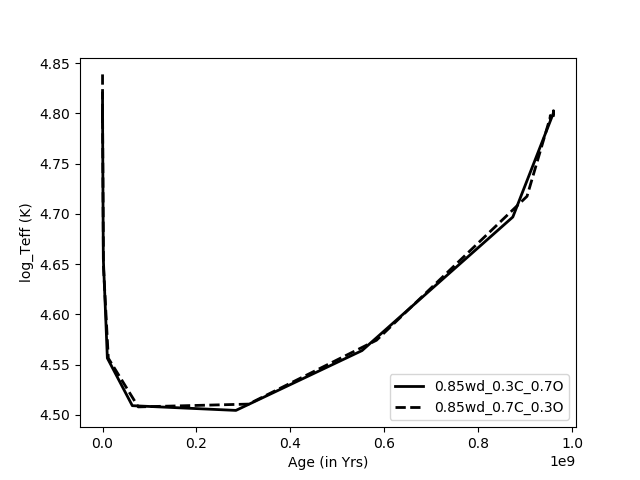} 
\caption{Variation in effective temperature $T_{eff}$ during the accretion phase of helium accreting white dwarf having 75000 K initial effective temperature.}
\label{fig:6}
\end{figure}


\section{Conclusion}
\label{sec:conclusion}
In the present work, we have studied the effect of slow accretion of helium-rich matter onto a WD of mass $0.85$ $M_{\odot}$. Carbon-Oxygen WDs with different compositions and different effective temperatures have been considered in our study. The accretion rate has been fixed to $5 \times 10^{-10}$ $M_{\odot}$ $yr^{-1}$, and the MESA stellar evolution code 
has been used for the simulations. The first two parameters we have studied are the surface gravity and radius of the WD. 
Independent of the composition or effective temperature, the surface gravity ($g$) increases and the radius of WD decreases monotonically as seen in Fig~\ref{fig:1} and \ref{fig:2}. At the end of each run, a sudden rise in radius and a sharp drop in $g$ is observed, which indicates the ignition of helium in the accreted layer. This result is independent of the effective temperature of the WD. The ignition of Helium towards the end of the run has been confirmed by studying the helium luminosity in Fig~\ref{fig:3} and \ref{fig:4}. For both the hot and cold WD, it is seen that helium luminosity increases rapidly at the end and crosses the threshold value to consider the ignition. 

It is interesting to note that there is a small difference in the radii of the Carbon dominated and Oxygen dominated WDs. 
Along with the electron degeneracy pressure, the outward pressure inside the WD has a small contribution from the ions as well which depends on the mass number of the ions. Thus in the equilibrium the total pressure,  $P = P_e +N_A kT/<A>$, balances the gravity. The ion pressure in oxygen dominated WD will be small due to the larger mass number of oxygen and hence will settle at a slightly smaller radius in equilibrium. This effect can be confirmed from the bottom panel of Fig~\ref{fig:1} and \ref{fig:2}. 

The next important result of our study is about the onset of helium ignition. A careful investigation of Fig~\ref{fig:1} and \ref{fig:2} shows that helium ignition starts slightly late if the carbon fraction is small in both the cold and hot types of WDs. Thus the WD with a smaller carbon fraction accretes more helium mass before ignition. As per our study, the dependence of helium ignition onset on carbon is mild but consistent. 

As discussed earlier the oxygen dominated WD has slightly smaller radius which corresponds to higher density. Higher density leads to higher electron degeneracy pressure for oxygen dominated WD. Thus, is order to compress the accreted material to ignite helium fusion slightly more mass accumulation would be required. This is the probable reason that helium ignition is slightly delayed in WD with (0.3C, 0.7O). 

The final outcome of accreting white dwarf of different abundances with slow accretion rate ($5 \times 10^{-10}$ $M_{\odot}$ $yr^{-1}$) results in single helium detonation at the base of accreted helium, which is in agreement with previous studies (Nomoto 1982; Piersanti {\em et al.} 2014). Finally, the effective temperature of the WD has also been studied during the accretion. It is seen that temperature initially decreases rapidly. Nevertheless, this happens only for a short duration, and after that, the temperature increases for both cold and hot WD.

\section*{Acknowledgements}
\vspace{-0.5em}
SG thanks SERB for financial assistance (EMR/2017/003714).

\end{document}